\documentclass
[hypertex,twocolumn,showpacs,preprintnumbers,showkeys]
{revtex4}

\usepackage{bm}
\usepackage{hyperref}
\usepackage{pictex}
\usepackage{graphicx}

\newcommand{\fr}[2]{{\textstyle \frac{#1}{#2}}}
\newcommand{\deltap}{(2\pi)^d\delta({\textstyle \sum\bm{p}_i)}}

\begin{document}

\title{Diagrammatics of the Dimensionally Reduced Action in
$\phi^4$ theory}

\author{C. Villavicencio}
\email{cristian@if.ufrj.br}
\author{C. A. A. de Carvalho}
\email{aragao@if.ufrj.br}%
\affiliation{%
Instituto de F\'{\i}sica, UFRJ, C.P. 68528, Rio de Janeiro, RJ 
21945-970, Brasil}%

\begin{abstract}
\noindent
We describe the Feynman rules for the construction of the \emph{
Dimensionally Reduced  Action} (DRA) for a $\phi^4$ field 
theory, as well as rules for the construction of the connected Green 
functions and the \emph{Dimensionally
Reduced Effective Action} (DREA), the effective action obtained
from the DRA. 
Two and four-point 
connected Green functions are calculated to illustrate the application of those
rules, and to exhibit the subtraction of UV divergences.
As the DRA formalism expands about a background 
field, 
instead of the high temperature expansion obtained in the usual dimensional 
reduced field theory, we derive an expansion that is defined for all range of
temperatures. 
We make a low temperature analysis of the DREA by 
considering soft modes. 
In particular, we study the case of near zero mass.
\end{abstract}



\maketitle

\noindent
Dimensional reduction
\cite{Weinberg:1980wa,Ginsparg:1980ef,Appelquist:1981vg,Nadkarni:1982kb} is a
powerful tool to describe thermal field theory for high temperatures,
reducing it to a stationary field theory which simplifies the
calculations, in particular for the description of relativistic heavy
ion collision experiments. 
It also describes very well the behavior of
high temperature semi-classical systems and critical phenomena
\cite{Zinn-Justin:2000dr}. 
Nevertheless, expanding around
infinite temperature introduces problems such as the appearance of spurious
divergences. 
Another problem is that the validity of the theory is restricted to very high
temperature systems, being the low-temperature regime not inaccessible
from this point of view.

The dimensionally reduced action (DRA), on the other hand, is
defined, and  well-behaved, for all ranges of temperature, and is free
of the spurious divergences that appear in the dimensionally reduced field
theory. 
It is constructed by incorporating quantum fluctuations in an expansion 
around a classical field configuration which satisfies the equations of motion
and the KMS conditions
\cite{Kubo:1957mj,Martin:1959jp}.
The DRA was proposed and developed in \cite{deCarvalho:2001xv} in $d$ spatial
dimensions for scalar and gauge fields, and studied in \cite{deCarvalho:2002tg}
for $d=0$, which corresponds to the quantum mechanical case.

The first objective in this article is to find simplified rules for the
computation of connected and one-particle irreducible (1PI) diagrams, which are
the physical thermal-correlators obtained in perturbation theory.
Although the rules we give here are for a $\phi^4$ theory, it is easy to
generalize for any kind of potential, and for other scalar field theories
involving more than one interacting field. 
The DRA for a fermionic field theory will be pending.
The second objective is the construction of the effective action from
the reduced action, or the
\emph{Dimensionally Reduced Effective Action} (DREA),  and its
analysis for soft modes and low temperatures.

This article is organized as follows: In section \ref{sec.DRA} we will
give a brief description of the DRA and the Feynman rules to construct
it;  in section \ref{sect-conn}, we will generalize the rules to
construct connected diagrams, and exhibit two examples, the self-energy
and the four-point connected diagrams at one-loop;
finally, in section \ref{sect-ea}, we will construct the DREA, and
generalize the rules for the construction of $1PI$ diagrams. 
Then, an analysis
of the effective action for soft modes and low temperatures will be 
made for the
massive case, and for small masses.

\section{Brief introduction to DRA}\label{sec.DRA}

\noindent
Following the construction described in \cite{deCarvalho:2001xv}, we will start
with the
definition of the DRA from the partition function ${\cal
Z}=\textrm{Tr}~\rho$, with the diagonal density matrix defined as 
$$
\rho[\phi,\phi]=\oint D\varphi~ \textrm{e}^{-S[\varphi]}.
$$
The integral $\oint$ is to be performed over all fields
$\varphi(\tau;\bm{x})$ that satisfy the boundary conditions in the
euclidean time $\varphi(0;\bm{x})=\varphi(\beta;\bm{x})=\phi(\bm{x})$
with $\beta=1/T$. 
The original action $S=S_F+S_I$ is the Klein-Gordon action in euclidean space,
where the free part is defined as
\begin{equation}
S_F[\varphi] = 
\int_{\tau 
{\bm{x}}}\frac{1}{2}\big[(\partial_\tau\varphi)^2+(\bm{\nabla}
\varphi)^2+m^2\varphi^2\big].
\label{S_F}
\end{equation}
Here, we use $\int_\tau\equiv\int_0^\beta  d\tau$
and $\int_{\bm{x}}\equiv\int d^dx$. 
Expanding the field $\varphi$ around a background field $\hat\varphi$
that satisfies the boundary conditions and the equations of motion for
the free theory,  we obtain the density matrix
$\rho=\exp\{-S_\textrm{\tiny RED} \}$, where $S_{\textrm{\tiny RED}}$
is the reduced action, defined as
$$
\textrm{e}^{-S_{\textrm{\tiny RED}}[\phi]}=
\textrm{e}^{-S_F[\hat\varphi]}\oint D\eta
~\textrm{e}^{-S_F[\eta]-S_I[\hat\varphi+\eta]},
$$
and where $\varphi=\hat\varphi+\eta$. 
Now,
the field integration is performed over all fields $\eta(\tau;\bm{x})$
that satisfy the boundary conditions
$\eta(0;\bm{x})=\eta(\beta;\bm{x})=0$. The background field is defined
as  
$$
\hat\varphi(\tau;\bm{x})=\int_{\bm{p}}
h(\tau;p)\tilde\phi(\bm{p})\textrm{e}^{i\bm{p\cdot}\bm{x}},
$$
where $\tilde\phi(\bm{p})$ is the Fourier transformation of
$\phi(\bm{x})$, and
\begin{equation} 
h(\tau;p)=\frac{\cosh[\omega_{{p}}(\tau-\beta/2)]}{\cosh(\omega_{p}
\beta/2)}.
\label{h}
\end{equation}
The integral in Fourier space is defined as
$\int_{\bm{p}}\equiv\int\frac{d^dp}{(2\pi)^d}$, and
$\omega_p=\sqrt{p^2+m^2}$. 
This background field then satisfies the
equations of motion and the boundary conditions
$\hat\varphi(0,\bm{x})=\hat\varphi(\beta,\bm{x})=\phi(\bm{x})$.  

In the construction and applications of the DRA, there will appear
three propagators. The first one is the $\phi$ propagator
\begin{equation}
G_\phi(p) = \frac{1}{2\omega_{p}\tanh
  (\beta\omega_{p}/2)},
\label{G_ph}
\end{equation}
which
is the inverse of the free action in (\ref{S_F}), in momentum space.
The second one is the free propagator of the $\eta$ fields
\begin{eqnarray}
 G_\eta(\tau_1,\tau_2;p)=\hspace{5.5cm}&&\nonumber\\
\theta(\tau_1-\tau_2)
\frac{\sinh(\omega_p\tau_2)\sinh[\omega_p(\beta-\tau_1)]}
 {\omega_p\sinh(\beta\omega_p)}  +\tau_1\leftrightarrow\tau_2,&&
\label{G_eta}
\end{eqnarray}
which satisfies the boundary conditions.
There is also a third propagator, $G_{th}$, constructed with both
propagators $G_\eta$
and $G_\phi$
\begin{eqnarray}
&& G_{th}(\tau_1-\tau_2;p)=h(|\tau_1-\tau_2|;p)G_\phi(p) \nonumber\\
&& \quad\equiv  G_\eta(\tau_1,\tau_2;p)
+h(\tau_1;p)h(\tau_2;p)G_\phi(p). \hspace{1cm}
\label{G_th}
\end{eqnarray}
In fact, we will see later that if we sum all the contributions for
$\phi$
correlation diagrams calculated with the DRA, the last relation will
appear in
all internal lines for connected diagrams, and in all loops for 1PI
diagrams.
This propagator is the Fourier transformation of the usual
Matsubara propagator:
$G_{th}(\tau;\bm{p})=T\sum_{n}\textrm{e}^{
\tau\omega_n }(\omega_n^2+\omega_{\bm{p}}^2)^{-1}$.

\begin{figure}
\centering
\includegraphics[scale=1]{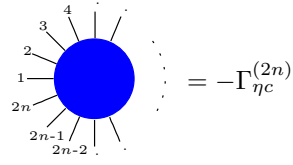}
\caption{ Diagrammatic representation of a DRA effective vertex with
$2n$
external $\phi$-lines.}
\label{DRA-diag}
\end{figure}

Consider the interaction part of the action as
\begin{equation}
S_I[\varphi]=\int_{\tau \bm{x}}\left[\frac{1}{2}\delta_m\varphi^2
+\frac{1}{4!}\lambda_0\varphi^4\right],
\label{S_I}
\end{equation}
$\lambda_0$ being the bare coupling with the counterterm
$\delta_\lambda=\lambda_0-\lambda\sim\lambda^2$, and $\delta_m$ being
the mass counterterm, defined as $\delta_m= m_0^2-m^2\sim\lambda_0$
$m_0$ being the bare mass. 
We will derive the rules to construct
the DRA, and calculate correlation functions for that
interaction, but our procedure can be easily generalized to any kind
of interaction term
described by a polynomial in powers of $\phi$.
After integration over the $\eta$ fields, hereafter called thermal
ghosts, 
re-exponentiating and neglecting constant terms independent of the
temperature,
the dimensional reduced effective action DRA will be then
$$
S_{\textrm{\tiny RED}}[\phi]
=S_{\textrm{\tiny RED}}[0]
+S_{\textrm{\tiny RED}}^F[\phi]
+S_{\textrm{\tiny RED}}^I[\phi],
$$
where $S_{\textrm{\tiny RED}}[0]$ is the vaccum contribution,
$S_{\textrm{\tiny RED}}^F[\phi]=S_F[\hat\varphi]$ is the quadratic
free part, and
$S_{\textrm{\tiny RED}}^I[\phi]$ the interactive part, defined
respectively as
\begin{eqnarray*}
S_\textrm{\tiny RED}[0] &=& \frac{1}{2} \ln\det
G_\eta^{-1}+\Gamma_{\eta c}^{(0)},\\
S_\textrm{\tiny RED}^F[\phi] &=&
\frac{1}{2}\int_{\bm{p}}G_\phi(p)^{-1}
\tilde\phi(-\bm{p})\tilde\phi(\bm{p}),\\
S_\textrm{\tiny RED}^I[\phi] &=& \sum_{n=1}^\infty
\frac{1}{(2n)!}\int_{\bm{p}_1\dots\bm{p}_{2n}}\Gamma_{\eta
c}^{(2n)}\tilde\phi_1\dots\tilde\phi_{2n},
\end{eqnarray*}
where $\tilde\phi_i=\tilde\phi(\bm{p}_i)$, and
the terms
$\Gamma_{\eta c}^{(2n)}(\bm{p}_1,\dots,\bm{p}_{n})$ are
defined as the \emph{effective vertices}, expressed diagrammatically
in
FIG.
\ref{DRA-diag}.  
The effective vertices are the terms resulting from the integration of
the $\eta$-field, and the sub-index $\eta c$ denotes that they are
constructed through connected $\eta$-diagrams. 
Then, they can be expressed as a series in powers of the coupling
constant 
\begin{equation}
\Gamma^{(2n)}_{\eta c}=\sum_{r\geq
1}\Gamma_{\eta c}^{(2n,r)}, \quad \mathrm{with}\quad
\Gamma_{\eta c}^{(2n,r)}\sim {\lambda_0}^r
\label{seriesGhc_2n-r}
\end{equation}
defined as
\begin{eqnarray*}
-\Gamma_{\eta c}^{(2n,r)} &=& \left.
\frac{(-1)^{r}}{r!}\frac{\delta^{2n} \langle 0|T_\eta
S_I[\hat\varphi+\eta]^r|0\rangle_{\eta
c}}{\delta\tilde\phi(\bm{p}_1)\dots
\delta\tilde\phi(\bm{p}_{2n})}\right|_{\phi=0},
\end{eqnarray*}
where the term  $\langle 0|T_\eta ~{\cal O}|0\rangle_{\eta c}$ has
the time ordering applied only to the $\eta$ fields, considering only
connected diagrams.
The effective vertices $\Gamma_{\eta c}^{(2n,r)}$ are totally
symmetric with respect to the momentum parameter.

Now, the DRA is expressed in powers of the coupling constant. 
If we are interested in calculating correlation functions up to order
$\lambda^R$, we just need to construct the DRA up to such order
$S_{\textrm{\tiny RED}}=\sum_{r=0}^R{\lambda_0}^{r}S_{\textrm{\tiny
red}}^{(r)}$.
After obtaining the DRA, we can then define the generating functional
for the $\phi$ fields, by adding a current source
\begin{equation}
{\cal Z}[J]  
=
\int D\phi ~\textrm{exp}\left\{-S_{\textrm{\tiny RED}}[\phi]
+\int_{\bm{x}}J(\bm{x})\phi(\bm{x})\right\}
\label{Z[J]}
\end{equation}
In the case $J=0$ we obtain the thermodynamic potential: 
$ {\cal Z}[0] =\textrm{exp}\left\{-\beta\Omega\right\}$.

\subsection{Feynman rules for the construction of
DRA}\label{sec.DRA_rules}

\begin{figure}
\centering
\includegraphics{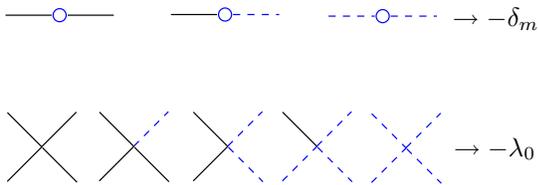}
\caption{Vertices from the interaction part of the Lagrangian
separated in the
background field part, containing the $\hat\varphi$ field (continuous line),
and the
fluctuation field part, containing $\eta$ (dashed line).}
\label{V_ij}
\end{figure}

\noindent
To obtain the effective vertices of the DRA, let us note first that due
to the
splitting of the original field into a background field and the
quantum
fluctuation, $S_I[\hat\varphi+\eta]$, we will obtain 8 diagrams
instead of two,
if we consider the interaction in (\ref{S_I}): Three vertices with two
legs
($-\delta_m$), and five vertices with four legs ($-\lambda_0$). FIG.
\ref{V_ij} 
shows diagrammatically the different vertices for the construction of
the DRA
effective vertices.

In terms of Feynman diagrams, the effective vertices are defined as
\begin{eqnarray}
 \Gamma^{(2n,r)}_{\eta c}(\bm{p}_1,\dots,\bm{p}_{2n})
=\frac{1}{(2n)!}\Big[\gamma_{\eta c}^{(2n,r)}(\bm{p}_1;\dots
;\bm{p}_{2n})&&\nonumber\\
+\mathrm{all}~\bm{p}_i\mathrm{-permutations}\Big]
 \deltap  &&
\label{def_Gamma_eta_c}
\end{eqnarray}
where $\gamma^{(2n,r)}_{\eta c}$ are the \emph{effective
vertex components}, the respective Feynman diagrams
to order ${\lambda_0}^{r}$ computed by integrating over  thermal
ghosts with $2n $  external $\phi$ legs. 
In general, we will use semicolons for nonsymmetric dependence on
momenta, and commas whenever quantities depend symmetrically. 
To construct the effective vertices, the Feynman rules are  almost the
usual
ones, except for small modifications.
\bigskip

\noindent
Rules to calculate the $2n$-points $\sim{\lambda_0}^r$ effective
vertex
 $-\Gamma_{\eta c}^{(2n,r)}$  in
 momentum space:
\begin{itemize}
\item Draw all the topologically distinct diagrams  with $2n$
external $\phi$-lines and connected internal $\eta$-lines formed with
$r$
vertices.
\item Multiply by $(-\delta_m)^c (-\lambda_0)^{r-c}/s $ for a 
diagram constructed with $c$ counterterm vertices, where $s$ is  the
symmetry factor.
\item Multiply by $h(\tau_i,\bm{p}_i)$, defined in (\ref{h}), for
every external $\phi_i$ line. 
\item Join the $\eta$ internal lines with the $G_\eta$ propagator,
defined in (\ref{G_eta}), imposing momentum conservation.
\item Integrate over euclidean time $\tau_i$, and  over undetermined
loop momenta.
\item Sum over all external momentum $\bm{p}_i$- permutations, and
multiply by $\deltap/(2n)!$.
\end{itemize}
\bigskip

A direct consequence of the construction of the effective vertices is
the fact that higher-point ($n>2$) effective vertices require more than one
vertex of order $\lambda_0$ for their construction. 
Specifically, the tree level $2n$-effective vertex (i.e. without
$\eta$-loops) will be of order ${\lambda_0}^{n-1}$. 
This will yield the useful relation
\begin{equation}
\Gamma_{\eta c}^{(2n)}\sim {\lambda_0}^{R\geq n-1}.
\label{r geq n-1}
\end{equation}
If we need to make corrections up to order $\lambda ^R$, we need the
DRA up to this order, which includes the tree level component
$\phi^{2(R+1)}$.
 
For example, $\lambda^2$ corrections, as we will see in section
\ref{G-4}, will include the term $\sim \Gamma_{\eta
c}^{(6,2)}\tilde\phi^6$, which is negative, and therefore makes the
functional integral divergent, as in the case found in
\cite{deCarvalho:2002tg}. 
Nevertheless, we can always include the next terms: in the present
case, if we go to order $\lambda^3$, we will include the positive
tree-contribution $\sim\Gamma_{\eta c}^{(8,3)}\tilde\phi^8$, giving
the correct convergence of the functional integral. 

This  procedure can be easily extended to the case of more than one
field involved. Particularly, in the case of gauge fields, the DRA
can be obtained by using the Feynman gauge, obtaining a tensorial
$\eta$-propagator
${G_\eta}^{\mu\nu}=\delta^{\mu\nu}G_\eta$.

In the next sections, we will exhibit two examples of one-loop
diagrams, and how to apply our Feynman rules. 
Moreover, we will derive more simplified rules to calculate connected
and 1PI diagrams.

\section{Connected diagrams, radiative corrections, and 
re\-normalization }\label{sect-conn}

\noindent
We already know the rules to construct the DRA. 
Depending on the order of the corrections to correlators or
observable quantities that we  want to obtain, we will construct it up
to that order. 
However, it is not necessary to construct the whole DRA, as we will show.
Here, we will present two examples, the most common ones, which are the
one-loop correction to the propagator, and the four-point vertex. 
Such radiative corrections will have ultraviolet divergences that will
be canceled by the two counterterms, $\delta_m$ and $\delta_\lambda$. 
In fact, those counterterms will be the only terms needed to remove UV
divergences to all orders, since the effective theory came from a
renormalizable theory. 

To calculate connected diagrams, we need the generating functional for
the connected green functions  $G_c[J]=\ln{\cal Z}[J]$ defined from
(\ref{Z[J]}). Expressed in terms of the $2n$ connected Green
functions, it is defined as
$$
G_c[J]=-\beta\Omega+
\sum_{n=1}\frac{1}{(2n)!}\int_{\bm{p}_1\dots\bm{p}_{2n}}G_c^{(2n)}
\tilde
J_1\dots \tilde J_{2n}  ,
$$
with $\tilde J_i\equiv \tilde J(\bm{p}_i)$, the
Fourier transformation of $J$. 
The connected Green functions $G_c^{(2n)}(\bm{p}_1,
\dots,\bm{p}_{2n})$ can be expressed in terms of a connected vertex
$\Gamma_c^{(2n)}$ with $2n$ external legs, which corresponds to the
amputated Green function
$$
G_c^{(2n)}
=-\Gamma_c^{(2n)}(\bm{p}_1,\dots,\bm{p}_{2n})G_\phi(p_1)\dots
G_\phi(p_{2n}),
$$
and the thermodynamic potential $\beta\Omega=\Gamma_c^{(0)}$. 
Like the DRA effective vertices, the connected vertices will be
totally symmetric, also following the relation in
(\ref{def_Gamma_eta_c}), and can be expressed as a series in the
renormalized coupling constant $\Gamma_c^{(2n)}=\sum_{r\geq
0}\Gamma_c^{(2n,r)}$, with $\Gamma_c^{(2n,r)}\sim\lambda^r$ as in
(\ref{seriesGhc_2n-r}).
Diagrammatically, $\Gamma_c^{(2n,r)}$
corresponds to all the Feynman diagrams of order $\lambda^r$ with
connected
$\eta$ and $\phi$ lines, with the special cases
\begin{eqnarray*}
\Gamma_c^{(0,0)}&=&\frac{1}{2}\ln \det G_\eta^{-1} +\frac{1}{2}\ln\det
G_\phi^{-1},\\
\Gamma_c^{(2,0)} &=&
-G_\phi(p_1)^{-1}(2\pi)^d\delta(\bm{p}_1+\bm{p}_2).
\end{eqnarray*}

\subsection{Self energy and mass renormalization.}\label{se}

\begin{figure}
 \begin{center}
 \includegraphics{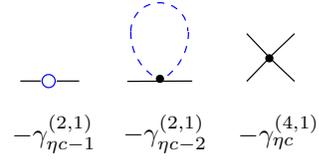}
 \end{center}
 \caption{Feynman diagram contributions to the effective vertices at
 order $\lambda_0$}
 \label{gamma-hc.O.lambda}
 \end{figure}

\begin{figure}
 \centering
 \includegraphics{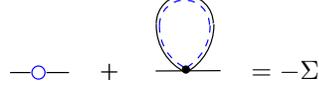}
 \caption{Diagram contribution to the self-energy. 
  The double line, solid and dashed, corresponds to the thermal
  propagator $G_{th}$.}
 \label{fig.g_c-2.1}
 \end{figure}

\noindent
Although this example was already discussed in
\cite{deCarvalho:2001xv}, we will explain it here in more detail, in
order to explain the notation involved.
Starting from the DRA, we need the two-point connected correlator
$\langle \tilde\phi_1\tilde\phi_2\rangle $ in
order to calculate the corrected $\phi$ propagator. To find the
self-energy we need only 1PI diagrams,
$$
G_c^{(2)}(\bm{p}_1,\bm{p}_2)
=\frac{(2\pi)^d\delta(\bm{p}_1+\bm{p}_2)}{G_\phi(p_1)^{-1}+\Sigma(p_1)
},
$$
where the self energy is defined as
$$\Sigma(p_1)(2\pi)^d\delta(\bm{p}_1+\bm{p}_2)=\Gamma^{(2)}_{
\phi\textrm{\tiny
    1PI}}(\bm{p}_1,\bm{p}_2),$$ 
which defines the one particle irreducible diagrams with respect to
the $\phi$ lines. 
In this case, the connected diagrams of order $\lambda_0$ are 1PI:
$\Gamma^{(2,1)}_c=\Gamma^{(2,1)}_{\phi\textrm{\tiny  1PI}}$. 
Since we want to compute corrections up to order $\lambda$, and
following the rules defined in Section \ref{sec.DRA_rules}, the
respective
Feynman diagrams needed to construct the effective vertices are 
 \begin{eqnarray}
\gamma^{(2,1)}_{{\eta c}-1} &=& \delta_m\int_\tau
 h(\tau,p_1)h(\tau,p_2),\\
\gamma^{(2,1)}_{{\eta c}-2}&=&
\frac{1}{2}\lambda_0\int_{\tau
\bm{k}}h(\tau,p_1)h(\tau,p_2)G_\eta
 (\tau,\tau;k),\\
 \gamma^{(4,1)}_{\eta c}
&=&
\lambda_0\int_{\tau
 }h(\tau,p_1)h(\tau,p_2)h(\tau,p_3)h(\tau,p_4).
 \label{ghc-4.1} 
 \end{eqnarray}
They are shown diagrammatically in FIG.~\ref{gamma-hc.O.lambda}. 
Now, we proceed to write the effective vertices as defined in
(\ref{def_Gamma_eta_c}). 
To calculate the connected vertex of order $\lambda$, we resort the
usual Feynman rules 
 \begin{eqnarray}
 \Gamma_c^{(2,1)}&=&
 \Gamma_{\eta c}^{(2,1)} +\frac{1}{2}\int_{\bm{k}} 
 \Gamma_{\eta c}^{(4,1)}(\bm{p}_1,\bm{p}_2,\bm{k},-\bm{k})G_\phi(k).
 \hspace{.8cm}
 \label{G_c-2.1}
 \end{eqnarray}
The sum of the contributions from the diagrams 
$\gamma_{\eta c-2}^{(2,1)}$ and  $\gamma_{\eta c}^{(4,1)}$ gives the
relation for the thermal propagator $G_{th}$ defined in (\ref{G_th}),
yielding the self energy
 \begin{eqnarray}
 \Sigma(\bm{p}) &=& 
 \int_\tau h(\tau,p)^2
 \left\{ \delta_m+\fr{1}{2}\lambda_0\int_{\bm{k}} G_{th}(0;k)\right\}.
 \hspace{.5cm}
 \label{sigma_1}
 \end{eqnarray}
Diagrammatically, the contributions to the self-energy are shown in
FIG.~\ref{fig.g_c-2.1}, where the sum of the thermal ghost line  and
the $\phi$ line  gives the thermal propagator, denoted as a double
line, solid and dashed.
The thermal propagator in (\ref{sigma_1}) can be written as 
 $$
 G_{th}(0;k)=\frac{1}{2\omega_k}(2n_B(\omega_k)+1),
 $$
where $n_B(\omega)=(\textrm{e}^{\beta\omega}-1)^{-1}$ is the
Bose-Einstein distribution. 
Then, we can identify the divergent part, which is independent of the
temperature. 
To remove it, we must set
 \begin{equation}
 \delta_m=-\lambda_0\int_{\bm{k}}\frac{1}{4\omega_k}.
 \label{deltam}
 \end{equation}
If we want to make higher corrections to the propagator, or one loop
corrections to the four-legged vertex, we need to consider the next
terms of the DRA.

Setting $\lambda_0=\lambda$, and integrating over euclidean time, the
corrected
propagator will be
 \begin{eqnarray*}
 G(p)^{-1} &=& G_\phi(p)^{-1}+\Sigma(p)\\
  &=& 2\omega_{p}\tanh(\omega_{p}\beta/2)\\
  && +\lambda\int_{\bm{k}}\frac{n_B(\omega_{k})}
   {2\omega_{k}}
   \frac{\beta\omega_{p} +\sinh(\beta\omega_{p})}
   {2\omega_{p}\cosh(\beta\omega_{p}/2)^2}\\
  &=& 2\omega\tanh(\omega\beta/2)
   +{\cal O}(\lambda^2),
 \end{eqnarray*}
where $\omega^2=p^2+m_D^2$ includes the Debye mass defined as
$G^{-1}(p^2=-m_D^2)=0$. 
In the present case, it corresponds to
$$
m_D^2=m^2+\lambda\int_{\bm{k}}\frac{n_B(\omega_{k})}
{2\omega_{k}},
$$
which is the well known result for the usual one-loop thermal mass
calculated with the real or imaginary time formalism.

\subsection{Four-point vertex correction and coupling constant
  renormalization } \label{G-4}

 \begin{figure}
 \centering
 \includegraphics{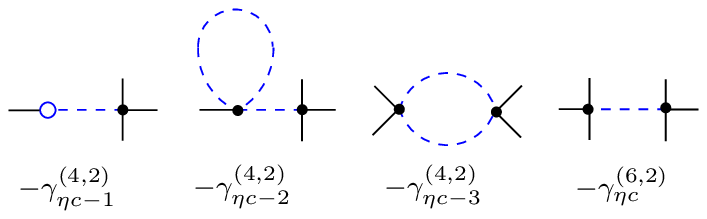}
 \caption{four and six leg contributions to the effective vertices of
order  ${\lambda_0}^2$}
 \label{g_4.2g_6.2}
\includegraphics{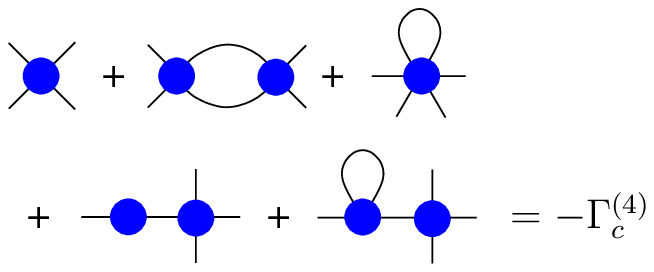}
 \caption{Relevant diagrams for the four-point connected vertex}
 \label{gamma_4c}
 \includegraphics{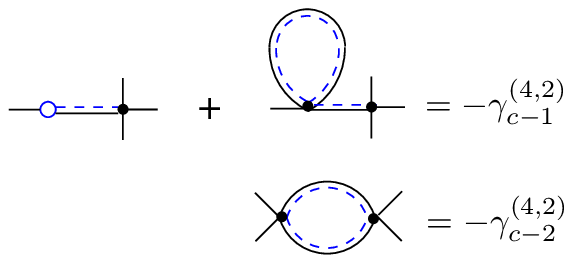}
 \caption{Contributions to the connected vertex of order
$\lambda^2$}
 \label{gc1-gc2}
 \end{figure}

For the four-point connected vertex,  we need the DRA up to order
$\lambda^2$. 
Yet, it is not necessary to calculate all the effective vertices. 
Apart from the order $\lambda_0$ contributions already calculated in
the previous example, we need the next contributions to the
four-point effective vertex, and the first contribution to the
six-point effective vertex, which are both of order ${\lambda_0 }^2$.
Their respective Feynman diagrams are shown in FIG.
\ref{g_4.2g_6.2}.
Having obtained the effective vertices, we proceed to calculate the
four-point connected vertices. 
The relevant Feynman diagrams are shown in \ref{gamma_4c}, and give as
a result
 \begin{eqnarray*}
-\Gamma^{(4)}_{c}
 &=&
\deltap\Bigg\{\gamma_c^{(2,1)}(\bm{p}_1,\bm{p}_2,\bm{p}_3,\bm{p}_4)
 \\
&&
+\frac{1}{4}\big[\gamma^{(4,2)}_{c-1}(\bm{p}_1;\bm{p}_2,\bm{p}_3,
\bm{p}_4)  
 +\bm{p}_1\leftrightarrow   
 \{\bm{p}_2,\bm{p}_3,\bm{p}_4\}\big]\\
 && +\frac{1}{3}\big[
 \gamma^{(4,2)}_{c-2}(\bm{p}_1,\bm{p}_2;\bm{p}_3,\bm{p}_4)
 +\bm{p}_2\leftrightarrow 
 \{\bm{p}_3,\bm{p}_4\}\big] 
 \Bigg\}
 \end{eqnarray*}
where $\gamma_c^{(4,1)}=\gamma_{\eta c}^{(4,1)}$, described in
(\ref{ghc-4.1}), and $\gamma_{c-1}^{(4,2)}$ and $\gamma_{c-2}^{(4,2)}$
are the the combination of the equivalent terms from the effective
vertices contracted with $\phi$ lines described in FIG.
\ref{gc1-gc2}.
Again, the combination of the $\eta$ and $\phi$ propagators gives the
thermal propagator $G_{th}$ in all the internal lines. In fact, this
will be a general rule in the construction of the connected
propagators.

Let us examine first the diagram $\gamma_{c-1}^{(4,2)}$. 
It is formed by two diagrams involving a closed loop and a
counterterm.  
The ultraviolet divergences are then simply removed by using the value
of the counterterm $\delta_m$ given in (\ref{deltam}). 
The result is
\begin{eqnarray}
  \gamma^{(4,2)}_{c-1}(\bm{p}_1;\bm{p}_2,\bm{p}_3,\bm{p}_4)
= \hspace{4cm}&&\nonumber\\
\quad
-2{\lambda_0}^2\int_{\tau_1\tau_2}h(\tau_1,p_1)h(\tau_2,p_2)h(\tau_2,
 p_3)h(\tau_2,p_4)&&\nonumber\\ 
 G_{th}(\tau_1-\tau_2;p_1)\int_{\bm{k}}\frac{n_B(\omega_k)}{
\omega_k } .&&
\label{gc1}
\end{eqnarray}

To find the divergent term in the diagram $\gamma_{c-2}^{(4,2)}$ is
not as simple as in the previous case. 
As one can see, it involves an external momentum dependent loop:
 \begin{eqnarray}
 \gamma^{(4,2)}_{c-2}(\bm{p}_1\bm{p}_2;\bm{p}_3,\bm{p}_4) = 
\hspace{4.1cm}&&\nonumber\\ 
-\frac{3}{2}{\lambda_0}^2 \int_{\tau_1\tau_2} h(\tau_1,p_1)
h(\tau_1,p_2)
 h(\tau_2,p_3) h(\tau_2,p_4)   \hspace{.1cm}&&\nonumber\\
 \int_{\bm{k}} G_{th} (\tau_1-\tau_2;k)G_{th}
 (\tau_1-\tau_2;|\bm{p}_1+\bm{p}_2-\bm{k}|).&&
 \end{eqnarray}

To identify the divergence, first we need to perform the euclidean
time integration. 
Since the counterterm is momentum independent, we just consider
$\bm{p}_i=0$
\begin{widetext}
\begin{eqnarray}
\gamma^{(4,2)}_c(0)  
 &=& 
 -{\lambda_0}^2
 \int_{\bm{k}}\left\{F(k) 
 +\frac{6\beta m+8\sinh(\beta m)+\sinh(2\beta
 m)}{16m\cosh(\beta
 m/2)^{4}}\frac{3}{32{\omega_k}^3}[2n_B(\omega_k)+1]\right\},
\label{g_4.2_c2_0}
\end{eqnarray}
with
\begin{eqnarray*}
F(k) &=& \frac{3 ~\mathrm{sech}(m\beta/2)^4}{16m^2{\omega_k}^2}\Bigg[
\frac{m^4}{k^4}\left(\frac{\sinh(\beta m)^2\cosh
(\beta\omega_k)}{\sinh(\beta\omega_k/2)^2}
-\frac{m}{\omega_k} 
\frac{ \sinh(2\beta m)}{\tanh(\beta\omega_k/2)}\right)\nonumber\\
&& \hspace{2.5cm} + \frac{m^3}{2\omega_kk^2}\frac{\beta m-\sinh(2\beta
m)}{2\tanh(\beta\omega_k/2)}
+\left(\frac{\beta m+\sinh(\beta
m)}{\sinh(\beta\omega_k/2)}\right)^2\Bigg].
\end{eqnarray*}
\end{widetext}

The function $F(k)$ is analytic in the real axis, and its integral is
finite for all temperatures.
The only divergent term in equation (\ref{g_4.2_c2_0}) is
$\int_{\bm{k}}\omega^{-3}$, in the case of $d=3$. 
Considering the fact that 
$$
\int_{\tau}h(\tau,p)^4=\frac{6\beta\omega_p+8\sinh (\beta\omega_p)
+\sinh
(2\beta\omega_p)}{16\omega_p\cosh(\beta\omega_p/2)^4},
$$
we can see that the divergent term is proportional to
$\gamma^{(4,1)}(0)$. 
Then, setting $\lambda_0=\lambda +\delta_\lambda$, with
$$
\delta_\lambda =\lambda^2
\frac{3}{32}\int_{\bm{k}}\frac{1}{\omega_k^3},
$$
we remove the divergence. 
We can now replace ${\lambda_0}^2=\lambda^2+{\cal O}(\lambda^3)$ in
all the expressions.

Note that in the case of $d\leq 2$ there are no divergent term since
the theory
is super-renormalizable. However, we will remove this term as a
general
prescription for any dimension.

There are no more counterterms. 
All the other divergences will be cancelled with $\delta_m$ and
$\delta_\lambda$, or will be divergences of higher order that can be
canceled by re-defining the counterterm with higher order terms.

\subsection{Feynman rules for connected diagrams}

\noindent
We can see in the last two examples that the thermal propagator is
present in all internal connected vertices. 
In fact, this will be a general rule, and to obtain connected Green
functions it is not necessary to calculate the effective vertices of
the DRA. 
Then, we just need two vertices to describe the different Feynman
diagrams: the four-point ($-\lambda_0$), and the two-point
($-\delta_m$).
\bigskip

\noindent
Rules to calculate the $2n$-points $\sim{\lambda_0}^r$ connected
vertex $-\Gamma_{c}^{(2n,r)}$  in momentum space:
\begin{itemize}
\item Draw all the topologically distinct connected diagrams
 with $2n$ external lines formed with $r$ vertices.
\item Multiply by $(-\delta_m)^c (-\lambda_0)^{r-c}/s $ for a diagram 
 constructed with $c$ counterterm vertices, where $s$ is the symmetry
 factor.
\item Multiply by $h(\tau_i,\bm{p}_i)$, defined in (\ref{h}), for
 every external $\phi_i$ line.
\item Join the internal lines with the $G_{th}$ propagator, defined
 in (\ref{G_th}), imposing momentum conservation.
\item Integrate over euclidean time $\tau_i$, and over undetermined
 loop momenta.
\item Sum over all external momentum $\bm{p}_i$- permutations and
 multiply by $\deltap/(2n)!$.
\end{itemize}

\section{The {\bf\emph{ Dimensionally Reduced Effective
Action}}}\label{sect-ea}

\noindent
We will construct the DREA using the standard method to obtain an effective action (see for example \cite{Amit:1984ms})
from the generating functional in equation (\ref{Z[J]})
\begin{equation}
a^{-1}\Gamma[\Phi]=-G_c[J]+\int_{\bm{x}}J(\bm{x})\Phi(\bm{x}),
\end{equation}
where $\Phi=\langle\phi\rangle_{J,T}$,  and $a$ is a small parameter
($a=\hbar$ in the semiclassical approximation). 
Now, let us redefine the fields, and scale the original action as 
$$ 
\phi=\Phi+\sqrt{a}\phi',\quad \eta=\sqrt{a}\eta',\quad
S[\varphi]\rightarrow a^{-1}S[\varphi].
$$

We will define $J=J_c+\sqrt{a}~\delta_J$, where $\delta_J$ will be a
counterterm which will remove tadpole diagrams, and $J_c$ is related
with the classical equation of motion through 
$\delta S_{\textrm{\tiny RED}}[\Phi]/\delta\Phi=J_c$. 
With these assumptions, the DREA will be
$$
\Gamma[\Phi]=S_{\textrm{\tiny RED}}[\Phi]-a\ln\int D\phi'\mathrm{e}^{
-S_{\textrm{\tiny RED}}^F[\phi']-\bar
S[\phi',\Phi]-\int_{\bm{x}}\delta_J\phi'},
$$
with
\begin{eqnarray}
\bar S[\phi',\Phi]&=&\sum_{n,r=1}^{\infty}\sum_{m=2}^{2n}
\frac{a^{m/2-1}}{(2n-m)!~m!}
\int_{\bm{p}_1\dots\bm{p}_{2n}}
\nonumber\\
&&\Gamma_{\eta c}^{(2n,r)} 
\tilde\Phi_{2n-m+1}\dots\tilde\Phi_{2n}
~\tilde\phi'_1\dots\tilde\phi'_m. 
\label{bar_S}
\end{eqnarray}
Now, we integrate over $\phi'$ using $G_\phi$ as the propagator. 
The resulting series will involve  N-particle irreducible diagrams. 
If we re-exponentiate again, the particle reducible diagrams will give
rise to tadpoles that will be cancelled by the counterterm $\delta_J$.
So, the effective action will be
\begin{eqnarray}
\Gamma[\Phi] &=& a\frac{1}{2}\ln\det G_\phi^{-1} +S_{\textrm{\tiny
RED}}[\Phi]\nonumber\\
&&-a\sum_{s=1}^{\infty}\frac{(-1)^{s}}{s!}\langle
0|S[\phi',\Phi]^s|0\rangle_{\phi\textrm{\tiny 1PI}}.
\label{Gamma-1}
\end{eqnarray}

Reorganizing in powers of the classical $\Phi$ field, the DREA will be
$$
\Gamma[\Phi]=\sum_{n=0}\frac{1}{(2n)!}\int_{\bm{p}_1\dots\bm{p}_{2n}}
\Gamma^{(2n)}\tilde\Phi_1\dots\tilde\Phi_{2n}.
$$
The vertices can be expressed in terms of number of loops 
\begin{equation}
\Gamma^{(2n)}=\sum_{l\geq 0}\Gamma^{(2n,l)} ,
\quad\mbox{ with}\quad
\Gamma^{(2n,l)}\sim\lambda^{n-1}(\lambda a)^l,
\label{G2n-l}
\end{equation}
where $l$ denotes the total number of loops, $\eta$ and
$\phi$, with the special cases
\begin{eqnarray*}
\Gamma^{(0,0)} &=& 0\\
\Gamma^{(0,1)} &=& \frac{a}{2}(\ln\det G_\phi^{-1}+\ln\det
G_\eta^{-1})\\
\Gamma^{(2,0)} &=&
G_\phi(\bm{p}_1)^{-1}(2\pi)^d\delta(\bm{p}_1+\bm{p}_2).
\end{eqnarray*}

To derive equation (\ref{G2n-l}), let us see first how the $a$
parameter changes the effective vertex  $\Gamma^{(2n,r)}_{\eta c}$
which involves $r$-vetices, $2n$ 
external $\phi$ legs, and $I_{\eta}$ internal $\eta'$ lines. 
Since the internal lines involve a pair of $\eta'$ fields they will
include a
term $a^{I_{\eta}}$, a power of $a^{-r}$ due to scaling of $S_I$,  and
one $a$ to get the common scaling factor in the full action. 
This mean
that $\Gamma^{(2n,r)}_{\eta c}\sim {\lambda_0}^ra^{I_{\eta}-r+1}$.
Using the well-known relation 
\begin{equation}
L=I-V+1,
\label{LIV}
\end{equation}
for diagrams with $L$-loops, $V$-vertices and
$I$-internal lines, we find then that $\Gamma^{(2n,r)}_{\eta c}\sim
{\lambda_0}^ra^{l_{\eta}}$. 

We would now like to express the $r$ parameter in terms of the
number of
$\eta'$-loops and external legs. 
Let us forget for a moment the diagrams which include the counterterm
$\delta_m$. Since the vertex contains four legs, the sum of the total
legs
($\eta'$ and $\phi$) will be $4r$, then we obtain the relation
$4r=2n+2I_{\eta}$.  
Together with relation (\ref{LIV}) we find that $r=l_\eta
+n-1$.
In the case of vertices with the counterterm $\delta_m$, they are
equivalent to a four-leg vertex with two legs joined by an internal
line. Then
$\delta_m\sim \lambda_0a$. 
The same will be done for the coupling constant where
$\delta_\lambda\sim\lambda^2a$.
The contribution to the effective action 
$S_{\textrm{\tiny RED}}[\phi]$ in (\ref{Gamma-1}) then will include 
the  effective vertices 
expressed in terms of number of $\eta$-loops
\begin{equation}
\Gamma_{\eta c}^{(2n,r)}=\Gamma_{\eta c}^{(2n,n-1+l_\eta)}
\sim {\lambda_0}^{n-1}(\lambda_0 a)^{l_\eta}
\label{G_hc-2}
\end{equation}

To find the other contributions from the integration of the $\phi'$
fields, let us consider a general $\phi$-1PI diagram
$\gamma_s^{(2n,l)}$, constructed with $s$ vertices.
From (\ref{bar_S}) and (\ref{Gamma-1}) we have that
$\gamma_s^{(2n,l)}\sim a\prod_{i=1}^{s}a^{m_i/2-1}\Gamma_{\eta
c}^{(2n_i,r_i)}$, where the sum of all external $\Phi$-legs and
internal $\phi$-lines gives
respectively 
$\sum_{i=1}^{s}(2n_i-m_i)=2n$ and 
$\sum_{i=1}^{s}2m_i=I_\phi$ (the total number of $\phi'$ internal
lines). With this, and equations (\ref{LIV}) and (\ref{G_hc-2}),
we obtain that $\gamma_s^{(2n,l)}\sim
{\lambda_0}^{n-1}(\lambda_0a)^{l_\phi+l_\eta} $ where $l_\phi+l_\eta=l$. Now, if
we sum all the
diagrams with $2n$ $\Phi$-legs and $l$-loops, and renormalize the mass
and the counterterms, we obtain  (\ref{G2n-l}).
 
\bigskip

The DREA will then be an infinite series in powers of the
fields, and in the small parameter $a\lambda$ raised to the number of loops. 
All powers of the fields will be accompanied by a coupling constant
$\gamma^{(2n)}\sim\lambda^{n-1}\Phi^{2n}$, independently of the number
of loops. 
Then, if $\Lambda $ defines the scale of validity of the theory, the
condition for convergence of the series is that
$\underline\Phi^2<\underline\lambda^{-1}$, where the underline stands
for scaling with $\Lambda$. 
So as $\underline\lambda\ll 1$, the range of validity of the field for
the convergence of the series is wide but always finite.

\subsection{Feynman rules for the DREA vertices}

Differently from the case of connected diagrams, the thermal propagator
will appear in all closed loops in the effective action. 
To illustrate it, we show one-loop corrections to the two and four-point vertices of the effective action.
Hereafter, we turn back to natural units ($a=1$).

In the case of two-point vertices, the one-loop correction is the
self-energy calculated in section \ref{se}, where the effective action
vertex is the inverse of the connected Green function
$\Gamma^{(2)}(\bm{p}_1,\bm{p}_2)=G_c^{(2)-1}(\bm{p}_1,\bm{p}_2)$.

\begin{figure}
\centering
\includegraphics{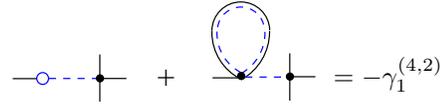}
\caption{One of the contributions to the four-point vertex of the effective
action.}
\label{gc1-g1}
\end{figure}

The correction to the four-point vertex is more illustrative. In
FIG.~\ref{gamma_4c}, the last two diagrams are two-particle reducible with
respect to the $\phi$ lines. The four-point vertex of the effective action then
will include only the first three diagrams at one loop. 
The result will change only the diagrams described by
$\gamma_{c-1}^{(4,2)}$, illustrated in FIG.~
\ref{gc1-gc2} and expressed in equation (\ref{gc1}) by cutting the
internal
$\phi$-line that joins the two DRA effective vertices, as can be seen
in FIG.
\ref{gc1-g1}. This gives as a result the replacement of the propagator
$G_{th}(\tau_1-\tau_2;p_1)$ by $G_\eta(\tau_1,\tau_2;p_1)$ in equation
(\ref{gc1}). 
So, as in the case of connected diagrams, the DREA
vertices will be constructed by considering only two vertices,
$-\delta_m$ and
$-\lambda_0$.

\bigskip
\noindent
Rules to calculate the $2n$-points and $l$-loops connected vertex
 $-\Gamma^{(2n,l)}$  in
 momentum space:
\begin{itemize}
\item Construct all the possible topologically distinct $l$-loops
connected diagrams
 with $2n$
external lines.
\item Multiply by $(-\delta_m)^c (-\lambda_0)^{r-c} $ for a  diagram 
constructed with $c$ counterterm vertices, and divide by the symmetry factor.
\item Multiply by $h(\tau_i,\bm{p}_i)$, defined in (\ref{h}), for
every external
$\phi_i$ line.
\item Join the internal closed loops with the $G_{th}$ propagator,
 defined in (\ref{G_th}), imposing momentum conservation.
\item Join the internal, non-loop lines with the $G_\eta$ propagator,
defined in (\ref{G_eta}).
\item Integrate over euclidean time $\tau_i$.
\item Integrate over undetermined loop momenta.
\item Sum over all external momentum $\bm{p}_i$- permutations, and
multiply by
$\deltap/(2n)!$.
\end{itemize}

\subsection{Soft modes in DREA}

\noindent
The main goal of the DRA and DREA is to obtain a reduced theory valid
for all ranges of temperature, in particular low temperatures, so this is the
main sector we would like to study.
In general, the macroscopic information obtained from microscopic systems is
encoded in their long range behavior.
Then, as we are interested in long distance behavior, we need mainly to extract 
low-momentum macroscopic parameters, considering only soft modes.
In the DREA, soft modes imply
the momentum expansion of the $\gamma $ terms, 
$$
\gamma^{(2n)}(\bm{p}_i)\approx\gamma^{(2n)}(0), \quad\mathrm{for}\quad
n>1,
$$
and
$$
\gamma^{(2)}(p)\approx\gamma^{(2)}(0)+\frac{\partial\gamma^{(2)}}{\partial
p^2}(0)p^2.
$$

Differently from the generator of connected diagrams, where all
temperature information is contained in the connected vertices, in
the effective action the classical fields $\Phi$, by definition, also
depend on temperature, so we can renormalize with a temperature
dependent factor $\Phi=\sqrt{Z}\Phi_R$. 
The effective action will then be
$$
\Gamma[\Phi_R]\approx\int_{\bm{x}}\left[\frac{1}{2}(\nabla\Phi_R)^2+\sum_{n=0}
\frac{g^{(2n)}}{(2n)!}\Phi_R^{2n}+{\cal O}(\nabla^4)\right],
$$
where the \emph{effective coupling constants} are defined as 
\begin{equation}
g^{(2n)}=Z^n\gamma^{(2n)}(0),
~~~\mathrm{with}~~~
Z^{-1}=\frac{\partial\gamma^{(2)}}{\partial p^2}(0).
\label{def-g2n}
\end{equation}

Usually, the scale of the theory may be taken as the cutoff for the
running mass defined as 
$$
\Lambda^{d-3}\lambda_{\textrm{\tiny RUN}}(p=\Lambda)\leq 1;
$$
perturbation theory is valid for momenta lower than this scale.
In general, soft modes are
defined as $p\ll m$.

Let us define the normalized effective
coupling constants for further analysis as
$$
g^{(2n)}(T;m;\lambda)=m^{2n-d(n-1)}\bar\lambda^{n-2}\bar g^{(2n)}(\bar
  T,\bar\lambda),
$$
where the bar on the parameters implies that they are scaled with
$m$: $\bar T=T/m$ and $\bar\lambda=\lambda m^{d-3}$. 
$\bar g$ can be expanded in the number of loops
$$
\bar g^{(2n)}(\bar T, \bar\lambda)=\sum_{l=0}\bar\lambda^l\bar
g^{(2n,l)}(\bar T).
$$

\begin{figure}
\centering
\includegraphics{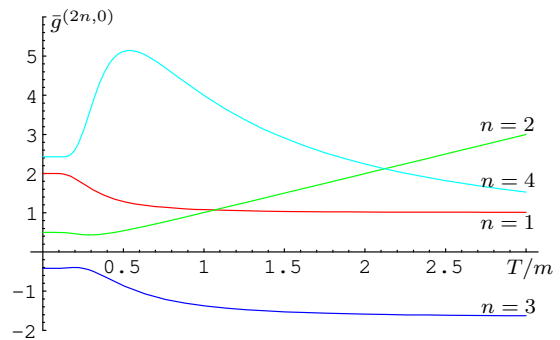}
\caption{Different two-leg effective couplings at tree-level for
$d=3$}
\label{graph_g-tree}
\end{figure}

FIG.~\ref{graph_g-tree} shows the evolution in temperature of the
different tree-level effective coupling constants up to $n=4$. 
We can see that, for temperatures higher than the mass parameter, the
effective couplings go smoothly to their known values from the usual
dimensional reduction formalism. 
Nevertheless, for low temperatures, their behavior changes abruptly. 
The reason for this change is the fact that the functions involved are
not analytical in $T$, but we can make expansions in
$\mathrm{e}^{-m/T}\ll 1$. 
If we investigate the high and low $T$ limits for the free
propagator, we find that
$G_\phi(T\rightarrow\infty)^{-1}=\beta\omega^2$ and
$G_\phi(T\rightarrow 0)^{-1}=\omega$. Expanding in soft modes without
renormalizing the fields, we obtain that the quadratic part of the effective
action will be 
$$
\Gamma[\Phi]_{\textrm{{\small free}}}
\left\{
\begin{array}{cl}
\overrightarrow{\mbox{\scriptsize $T\to\infty$}}
& \beta\int_{\bm{x}}\frac{1}{2}\Big[(\nabla\Phi)^2+m^2\Phi^2\Big]\\
\overrightarrow{\mbox{\scriptsize $T\to 0$}}
&\int_{\bm{x}}\Big[\frac{1}{2m}(\nabla\Phi)^2+m\Phi^2\Big]+{\cal
O}(\nabla^4),
\end{array}\right.
$$

\subsection{Soft modes, low temperatures, and small masses}

\noindent
We know that if we start from a massless theory, the thermal bath will
provide a temperature dependent mass of the order of the coupling
constant, after radiative corrections.
Resummation techniques \cite{Pisarski:1988vd,Braaten:1989mz} include
these thermal masses, in order to remove infrared divergences.
Other formalisms use small masses in order to regulate the infrared
divergences. For example, by starting with a massless theory, then adding
a mass term of the order of the perturbation parameter, and
subtracting it as part of the interaction Lagrangian
\cite{Frenkel:1992az,Arnold:1994ps}.

Consider the scale of the theory $\Lambda$, such that
$\underline m^2\sim\underline\lambda<1$, where the underline implies scaling with $\Lambda$: $\underline
m^2=m^2/\Lambda^2$ and 
$\underline\lambda\equiv\lambda\Lambda^{d-3}$.
From dimensional analysis, the different vertex contributions will have the
form
\begin{equation}
\gamma^{(2n)}(\bm{p}_i;T;m;\lambda)=
m^{n-d(n-1)}\bar\lambda^{n-1}
\bar\gamma^{(2n)}(\bm{\bar p}_i;\bar T;\bar\lambda),
\label{bar-gamma-2n}
\end{equation}
which can be expanded in the number of loops as
\begin{equation}
\bar\gamma^{(2n)}(\bm{\bar p}_i;\bar T;\bar\lambda)=
\sum_{l=0}\bar\lambda^l\bar\gamma^{(2n,l)}(\bm{\bar p}_i;\bar T).
\label{bar-gamma-2nl}
\end{equation}
We want to know of which order in $\underline\lambda$ are the
vertices. As we are interested in soft modes and low temperatures, we
must consider them of the order of the mass parameter. 

First note that $\bar\lambda\sim\underline\lambda^{(d-1)/2}$. This
implies that for low masses it is not possible to use perturbation
theory for $d=1$, as we can see from equation (\ref{bar-gamma-2nl}). 
Scaling with $\Lambda$ the external factors in equation
(\ref{bar-gamma-2n}), we obtain
$$
\gamma^{(2n)}(\bm{p}_i;T;m;\lambda)
\sim \underline\lambda^{1/2}\bar\gamma^{(2n)}(\bm{\bar p}_i;\bar
T;\bar \lambda)
\Lambda^{n-d(n-1)}.
$$
The range of convergence of the fields must be
$\underline\Phi^2<1$, from the definition of the cutoff.
That is not the case if we renormalize the fields as in equation
(\ref{def-g2n}).
Expanding in soft modes, considering the external momenta $|\bm{p}|\ll
m$ we obtain
$$
g^{(2n)}(T,m,\lambda)\sim \underline\lambda^{(n+1/2)}\bar g^{(2n)}(\bar
  T,\bar \lambda)\Lambda^{2n-d(n-1)},
$$
so in this case we need, for convergence, that
$\underline\Phi_R^2<\underline\lambda^{-1/2}$.

\begin{figure}
\centering
\includegraphics{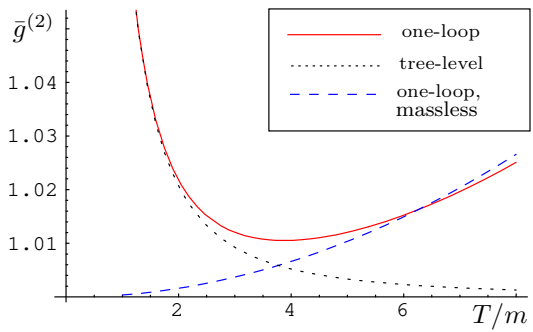}
\caption{Two-leg effective coupling constant in $d=3$ considering  a
small
mass $m^2=\lambda\Lambda^2$ at: tree-level (dotted line), one loop
(solid line),
and
one-loop treating the mass as a perturbation (dashed).}
\label{g2msmall}
\end{figure}

As an example, let us compute the one-loop corrected effective coupling
$g^{(2)}$ for $d=3$,  considering a small  mass
$m^2=\lambda\Lambda^{2}$ calculated in two different ways: one with
massive propagators, and the other by considering the mass as a
perturbation parameter in the interacting Lagrangian
${\cal L}_I=\frac{1}{2}m_0^2\varphi
^2+\frac{1}{4!}\lambda_0\varphi^4$. 
Following the procedures described for
the self-energy, and renormalizing the fields, we derive the result
which is shown
graphically in FIG.~\ref{g2msmall}. 
In the case of the one-loop corrected
effective coupling (solid line), calculated with the mass included,
for lower temperatures it behaves like
tree-level (dotted line), as already shown in FIG.
\ref{graph_g-tree}. 
As the
temperature starts to grow, it behaves like the one-loop corrected
effective
coupling calculated by considering the mass as a perturbative
parameter (dashed line).
Therefore, we cannot treat the mass as a
perturbation for low temperatures. This is because we have expanded in
soft
modes for momenta small compared with the mass. 
When we consider massless
propagators, the expansion in momenta must be made by comparing with
temperature, so,
for lower values of temperature, the expansion does not make sense
 as can be seen
in the
propagator, if we consider zero mass and temperature:
$G_\phi^{-1}\rightarrow p$.

\section{Conclusions}

We have generalized the rules for the construction of the connected
Green functions from the reduced theory, and the construction of the
DREA in a simple way.
From the two examples of one-loop connected propagators, we show that
it is easy to find the temperature independent divergences to be 
renormalized by the mass and the coupling constant. Also, we identify
the Debye mass, which corresponds to the usual thermal mass. 
The rules can easily be changed to other kinds of 
field theories.

Maybe the most interesting part of this work is the analysis of the 
effective action at low temperatures for soft modes. 
We show graphical examples of different effective coupling constants
which for low temperatures change from their extrapolated high
temperature behavior due to the 
non-analytic nature of the functions involved in the temperature.
A dimensional analysis of the effective couplings shows us that, for
low temperatures, perturbation theory is also valid for $d>1$ if we 
consider a small mass of the order of the coupling constant. In 
particular, soft modes, for near-zero temperatures, must be dealt with by including a mass term.

The DREA can, in principle, describe a 
great variety of low energy theories expanded about the classical 
limit.
The Hamiltonian thus obtained corresponds to a Ginsburg-Landau
coarse-grained free 
energy constructed from a microscopic theory. 

In a forthcoming publication, we will use the DREA to analyze a
Lagrangian with
a small negative mass-squared term, in order to describe a phase transition,
and extract critical parameters.

\begin{acknowledgments}
C.A.A. de C. acknowledges financial support from CAPES, CNPq,
FAPERJ and FUJB/UFRJ.\\
 C.V. acknowledges financial support from CNPq/CLAF. 
\end{acknowledgments}

\bibliography{drea}

\end{document}